# Analysis on image recovery for digital Fresnel hologram with aliased fringe generated from self-similarity of point spread function

**BYUNG GYU CHAE**[*]

*Biomedical Imaging Group, Electronics and Telecommunications Research Institute, 218 Gajeong-ro, Yuseong-gu, Daejeon 305-700, South Korea*
*\*bgchae@etri.re.kr*

**Abstract:** We analyze the aliasing phenomenon for digital Fresnel hologram with an enhanced numerical aperture (NA). The enhanced-NA digital hologram acquired computationally or optically at a closer distance from the object has an aliased fringe generated by undersampling process of the Fresnel prefactor. The point spread function known as Fresnel factor reveals a self-similar envelope when being sampled, which becomes a crucial mechanism in making this type of aliasing fringe of hologram. We describe that as the enhanced-NA hologram involves already the complementary aliased fringe that might come up in the reconstruction process, the robust recovery of object image can be realized. These behaviors are confirmed through numerical simulation. Based on the analysis of aliased hologram fringe, we provide a method for reconstructing the object image from the enhanced-NA digital Fresnel hologram without the shrinkage of image size.

*Keywords*: Digital Fresnel hologram; Holographic image; Numerical aperture; Point spread function

## 1. Introduction

Digital hologram obtained either computationally or optically in the in-line holographic system has a finite spatial-bandwidth product due to its sampling on a digitized device [1-3]. The sampling condition of this system to avoid an aliasing effect has been widely studied especially for the Fresnel diffraction fields [4-6]. The present device has a finite size and consists of pixel array with several micrometer pixel pitch, and thus, there exist some restrictions about an image size and resolving power of reconstructed image. Particularly, the resolution enhancement is very essential in improving the image quality, where the image resolution is known to be determined by the numerical aperture (NA) of hologram [7,8]. In holographic display, we have studied that the viewing-angle of a holographic image could be increased by synthesizing the digital hologram with a high NA [9]. In digital holography, several researches [10,11] such as a synthetic aperture imaging have been performed to obtain the restored image with a high resolution.

Generally, the NA of on-axis hologram is proportional to its lateral size and in inverse proportion to a reconstruction distance. We can expect that the high numerical aperture in digital hologram with a finite specification is accomplished by acquiring the hologram data at a distance close to the object. In this case, the aliasing error of hologram fringe could be produced, and hence, it is important to comprehend the effect of aliasing error on the hologram synthesis and acquisition processes. The analysis of aliasing phenomenon for the digital hologram is rather complicated despite of an abundant literature [4-6,9,12-14], because it is the sampling of a propagating diffraction field at a proper distance other than a simple Fourier analysis of image intensity. The diffracted Fresnel field is described as the convolution of the object field and point spread function known as optical kernel. The point spread function $h(x, y)$ has a quadratic phase term showing a rapid oscillation with lateral space [15]:

$$h(x, y) = \frac{e^{ikz}}{i\lambda z} \exp\left[i\frac{\pi}{\lambda z}(x^2 + y^2)\right], \qquad (1)$$

where $k$ is wavenumber of $\lambda/2\pi$ with wavelength $\lambda$, and $z$ is a distance between object plane and hologram plane. Since the Fresnel diffraction involves the Fourier transform of a product of object field and quadratic phase term, the sampling theorem based on the Fourier analysis is applicable by itself [5,6]. Typical approach to interpret an aliasing effect for diffracted field is to investigate the sampling condition about the quadratic phase term. In this analysis [12-14], the sampling condition is defined for the geometrical configuration such as the object field size and propagation distance.

In practical applications, when the digital hologram of real or imaginary value of diffraction field is loaded on the digitized device, an aliased fringe of hologram could be generated owing to a finite pixel pitch of device although it obeys the sampling condition with respect to object plane. This type of aliasing error can be analyzed by using a Fresnel prefactor in the hologram plane [13], where the sampling rate of diffracted field should fulfil the Nyquist criterion as well. When the pixel pitch of digital hologram is larger than sampling pitch of diffracted field, the undersampling of hologram fringe arises. The digital hologram generated at a close distance will reveal a severe undersampling phenomenon.

It has been known that optimal sampling condition is possible only at a specific distance [12-14], and thus in the reconstruction process of the digital hologram with a high NA, the hologram is undersampled while the restored image is oversampled. This situation obscures whether or not the image retrieval is well defined in real system. Furthermore, when the hologram fringe of a point source, known as a Fresnel zone plate is undersampled, the replica patterns are created [16,17]. It is not certain that the resolution of the restored images from these zones is simply interpreted by the NA of entire hologram [6,13]. However, we observed in optical experiments that the digital hologram with a severe aliasing error reconstructs the original image with a high resolution, and explained that the image resolution is determined by the extent of diffraction fringe of hologram aperture [9].

Recently, the fractal structure of diffraction pattern for various optical elements has been widely reported [18-20]. The fractal zone plates reveal a self-similar pattern of axial irradiance. Especially, even a diffraction wave from the linear binary grating shows the fractal structure of transverse beam profiles. The intrinsic character of point spread function acting as optical kernel in optical wave propagation may affect these behaviors. Thus, the analysis of the internal structure of optical kernel function helps us understand an aliased phenomenon of hologram fringe.

In this study, we analyze minutely an aliasing phenomenon of hologram fringe generated by undersampling process of the Fresnel prefactor, and then investigate the effect of aliasing error of hologram fringe on image recovery. It is elucidated that the point spread function has an internal structure of envelope curve when being sampled, which shows a self-similarity depending on a sampling pixel pitch. The fractal-like structure of optical kernel function becomes a key factor inducing an aliasing fringe of hologram. We carry out numerical simulation for the hologram having the enhanced numerical aperture to study this property.

## 2. Analysis on aliasing effect for digital Fresnel hologram with enhanced numerical aperture

*2.1 Aliasing effect of on-axis digital Fresnel hologram*

The aliasing interpretation should be differently applicable for various methods describing a diffraction wave. Only a conventional Fresnel transformation approach with a single Fourier transform makes it possible to interpret the image resolution [9], based on the Rayleigh criterion. Therefore, it is important that the stable recovery of a holographic image is consistent in this framework.

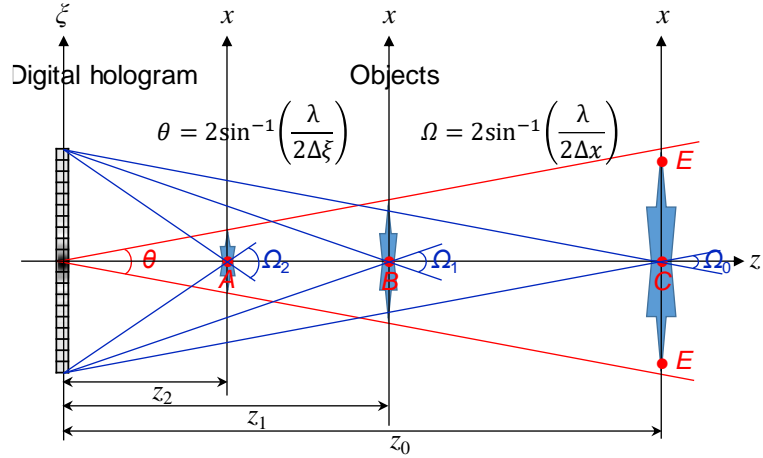

Fig. 1. Schematic diagram for the configuration of digital Fresnel hologram and objects in a hologram synthesis in the in-line holographic system. The object resolution increases at a closer distance from digital hologram. Red and blue lines indicate the diffraction zone with respect to pixel pitch of hologram and object, respectively.

We consider the computer-generated hologram in the in-line holography system, as depicted in Fig. 1. If the coaxial plane wave with a unit amplitude is used as a reference wave, the real or imaginary component of diffraction field becomes the hologram in the target plane. The Fresnel diffraction formula describing the Fresnel field $U(\xi,\eta)$ diffracted from the object field $U_0(x,y)$ can be written by

$$U(\xi,\eta) = \frac{e^{ikz}}{i\lambda z} \iint U_0(x,y) \exp\left\{\frac{i\pi}{\lambda z}\left[(\xi-x)^2 + (\eta-y)^2\right]\right\} dxdy. \qquad (2)$$

In order to investigate aliasing effect for diffraction field sampling in the conventional Fresnel transform, we expand above formula as follows,

$$U(\xi,\eta) = \frac{e^{ikz}}{i\lambda z} \exp\left[\frac{i\pi}{\lambda z}(\xi^2+\eta^2)\right] \boldsymbol{FT}\left\{U_0(x,y)\exp\left[\frac{i\pi}{\lambda z}(x^2+y^2)\right]\right\}. \qquad (3)$$

This equation is represented as a multiplication of two separate parts: a preceding point spread function and the Fourier transform $\boldsymbol{FT}$ of a product of object field and embedded point spread function. The spectrum analysis of propagating diffraction field is carried out by using the Fourier transform term.

For convenience, one-dimensional description for both fields discretized into $N \times N$ pixels will be used hereafter. According to the Fourier analysis, the spatial frequency is $\xi/\lambda z$ with lateral coordinate, and thus the pixel resolution $\Delta\xi$ of hologram field is defined in terms of resolution $\Delta x$ of the object field [1]:

$$\Delta\xi = \frac{\lambda z}{N\Delta x}. \qquad (4)$$

This value is explained by the NA of digital hologram, which in free space, is geometrically given by

$$\mathrm{NA} = \sin\Omega_{\mathrm{NA}} = \frac{N\Delta\xi}{2z}. \qquad (5)$$

Based on the Abbe criterion, the resolution limit $R_{Abbe} = \Delta x$ of restored image has the following relation [9,21],

$$R_{Abbe} = \frac{\lambda}{2\,NA}. \tag{6}$$

Figure 1 shows the configuration of on-axis digital Fresnel hologram and objects in the hologram synthesis. Considering a finite device with constant pixel pitch, the object size at a closer distance is reduced due to $\Delta x$ decrement in Eq. (4). This means that to realize the object image with a higher resolution, the digital hologram should be acquired at a closer distance from the object. Here, we set $z_0$ as the distance that both pixel sizes of object plane and hologram plane are the same, $z_0 = N\Delta x^2/\lambda$. The digital hologram synthesized at a distance lower than $z_0$ can retrieve the image with a resolution higher than the pixel pitch of digital device, and hence, for convenience we define this type of hologram as an enhanced-NA hologram. The diffraction angle $\Omega$ with respect to the pixel resolution of object field increases with decreasing a distance:

$$\Omega = 2\sin^{-1}\left(\frac{\lambda}{2\Delta x}\right). \tag{7}$$

Furthermore, the object field-views in Fig. 1 are confined in the diffraction zone by a diffraction angle $\theta$ of hologram pixel pitch to evade the high-order diffraction errors,

$$\theta = 2\sin^{-1}\left(\frac{\lambda}{2\Delta \xi}\right). \tag{8}$$

As described in previous Section, the aliasing phenomenon of digital hologram is examined in the manner of searching for the sampling condition of the quadratic phase term in parenthesis of **FT** [12-14]. It is assumed that the object field is a slowly varying function compared to the quadratic phase term. To avoid an aliasing effect in the digital hologram synthesis, the sampling rate $f_s$ of object field should be larger than two times the maximum spatial frequency $f_{x,\max}$:

$$f_s \geq 2 f_{x,\max} = \frac{2 x_{\max}}{\lambda z}, \tag{9}$$

where the maximum frequency $f_{x,\max}$ is estimated from the embedded quadratic phase $\phi(x,y)$ of Eq. (3):

$$f_{x,\max} = \frac{1}{2\pi}\left|\frac{\partial \phi(x,y)}{\partial x}\right|_{\max} = \frac{x_{\max}}{\lambda z}, \tag{10}$$

whose value depends on both a distance and object field size. If a sampling rate in the object plane is put to be $\Delta x^{-1}$ and the relation of object field size that $N\Delta x = 2|x_{\max}|$ is used, well-sampling condition with respect to a synthesis distance is given by

$$z \geq \frac{N\Delta x^2}{\lambda}. \tag{11}$$

On the other hand, as shown in Fig. 1, if the pixel pitch $\Delta \xi$ of hologram is larger than $\Delta x$ of the object field at a closer distance, the hologram fringe is undersampled by its finite pixel pitch of hologram plane. When the hologram is synthesized at a distance lower than $z_0$, the undersampling is inevitable although the diffracted field propagates in the well-defined area from Eq. (4) and it follows the sampling rule of Eq. (11). This undersampling process will form an aliased hologram fringe, which is unusual behavior compared to that of previous spectral analysis. This phenomenon is characterized by using a quadratic phase factor in front of **FT** operation in Eq. (3). Through similar derivation to Eqs. (9)-(11), we can obtain the following sampling condition in the hologram plane:

$$z \geq \frac{N \Delta \xi^2}{\lambda}. \tag{12}$$

By incorporating Eq. (4) to Eq. (12), the sampling operation without aliasing error is possible only at a constant distance [12,13],

$$z_c = \frac{N \Delta x^2}{\lambda}. \tag{13}$$

We know that in the hologram with enhanced numerical aperture, an aliasing error of hologram fringe could be generated from the undersampling of oscillating phase of a preceding Fresnel factor. Here, the diffraction spectrum related to **FT** term is well sampled because it places within the diffraction area by object pixel pitch in Fig. 1. Surely, the Riemann integral method can avoid this kind of aliased error because the pixel resolution of hologram plane can be arbitrary controlled. However, it is not realistic because actual device has a fixed pixel specification, as illustrated in Fig. 1. Either the convolutional method or angular spectrum method having a double Fourier transform can also evade this type aliased fringe, where the low-pass filtering takes place to prevent the aliasing error [9]. This approach is not under consideration, because it also deviates from our research direction trying to find the method for obtaining a high-resolution image.

*2.2 Stable recovery of holographic image under severe aliased hologram fringe*

The reconstruction process is a backward propagation from digital hologram to object plane in Fig. 1, where the diffraction equation is calculated through a reverse transform of Eq. (3):

$$U_0(x,y) = \frac{ie^{-ikz}}{\lambda z} \exp\left[-i\frac{\pi}{\lambda z}(x^2 + y^2)\right] \textbf{\textit{IFT}}\left\{U(\xi,\eta)\exp\left[-i\frac{\pi}{\lambda z}(\xi^2 + \eta^2)\right]\right\}. \tag{14}$$

Here, we note that both quadratic phase terms change places. The Fresnel prefactor of Eq. (3) is conversely embedded in the parenthesis of inverse Fourier transform **IFT**. The sampling condition derived from the quadratic phase term in the **IFT** operation appears to be the same as Eq. (12). The minimum distance to avoid an aliasing error appears to be $z_0$, and thus in the reconstruction process, the digital holograms synthesized at a distance lower than $z_0$ violate the sampling rule although these holograms are synthesized on the basis of well-defined sampling condition of Eq. (11). This means that considering its reconstruction process together, the well-sampling range seems to be limited to a constant distance $z_c$.

In the enhanced-NA digital hologram, the hologram fringe in the parenthesis of **IFT** has already complementary Fresnel factor with the aliasing error that might come up in the reconstruction process. Since the fringe error arises due to the increase of the maximum spatial frequency, an error extent of fringe might be described by a ratio of the maximum frequencies. When the extent of error occurred in the hologram synthesis at an arbitrary distance $z_n$ below $z_0$ is represented as

$$\text{Error}\left(\frac{z_0}{z_n}\right), \tag{15}$$

the error amount that happens in the reconstruction process will be the same as this value. Thus, even severe aliasing error generated from the hologram synthesis will be accurately compensated. The extent of Error() function can be estimated to be a portion of aliased fringe occupying the total fringe. From this, we find that the original image can be stably reconstructed despite of this type of aliasing error of hologram fringe.

*2.3 Matrix formalism for image recovery of on-axis digital Fresnel hologram*

We explain the property of image recovery through the matrix formalism. In Eq. (3), the column vector $\boldsymbol{\xi}$ of the diffraction field with respect to object vector $\mathbf{X}$ is represented as follows:

$$\boldsymbol{\xi} = \boldsymbol{\Psi}_{Fr}\mathbf{X}. \tag{16}$$

The Fresnel matrix $\boldsymbol{\Psi}_{Fr}$ is constructed by using the Fresnel factor and Fourier matrix. The component of Fresnel matrix is given by

$$\Psi_{mp} = H_\xi^m \omega^{mp} H_x^p, \tag{17}$$

where $\omega$ is $e^{-2\pi i/N}$ and the component of Fresnel factor $H_x^m$ is

$$\exp\left(\frac{i\pi}{\lambda z}m^2 \Delta x^2\right). \tag{18}$$

The Fresnel matrix has a unitary property because the transpose conjugate matrix is directly the inverse matrix [22]:

$$\sum_{p=0}^{N-1} \Psi_{mp}\Psi_{pm'}^* = \delta_{mm'}. \tag{19}$$

The original object can be completely reconstructed by inverse Fresnel transform when there is no aliasing effect:

$$\mathbf{X} = \boldsymbol{\Psi}_{Fr}^{-1}\boldsymbol{\xi}. \tag{20}$$

Furthermore, it is not hard to see that the Fresnel matrix periodically undersampled is unitary, and thus this undersampled hologram can recover original image, but modulated replica image appears due to the limited diffraction area by sampling pitch [5,6].

As previously stated, a conventional approach to extract the sampling condition is to deal with the embedded Fresnel factor $\mathbf{H}_x$. Whereas, in case of digital hologram synthesized at a closer distance lower than $z_0$ in Fig. 1, the undersampling of hologram fringe occurs only for the Fresnel prefactor $\mathbf{H}_\xi$, where the spectrum term $\boldsymbol{\omega}\mathbf{H}_x$ related to the Fourier transform in Eq. (3) is correctly sampled. If the aliased Fresnel prefactor is put to be $\mathbf{H}_{\xi,alias}$, the matrix form of Eq. (16) becomes

$$\boldsymbol{\xi} = \mathbf{H}_{\xi,alias}\boldsymbol{\omega}\mathbf{H}_x\mathbf{X}. \tag{21}$$

Since the aliased Fresnel factor that happens during inverse transform is like $\mathbf{H}_{\xi,alais}^{-1}$, the inverse Fresnel transform is written by

$$\mathbf{X} = \mathbf{H}_x^{-1}\boldsymbol{\omega}^{-1}\mathbf{H}_{\xi,alias}^{-1}\boldsymbol{\xi}. \tag{22}$$

Substituting the diffraction field of Eq. (21) into above equation, we can extract below identity with respect to Fresnel factor:

$$\mathbf{I} = \mathbf{H}_{\xi,alias}^{-1}\mathbf{H}_{\xi,alias}. \tag{23}$$

This type of aliased error from undersampling of Fresnel prefactor is canceled out, and hence, the original image is restored even at overly aliased hologram fringe.

Meanwhile, the fact that the original image is completely recovered at a proper distance does not imply the well-description of diffraction field propagation. This scheme should be able to obviously describe the propagating behavior of diffracted wave from the reconstructed image. We can suppose that the error amount that required at a distance $z$ away from the imaging plane is smaller than the quantity arisen at a proper distance $z_n$:

$$\text{Error}\left(\frac{z_0}{z}\right) \leq \text{Error}\left(\frac{z_0}{z_n}\right). \tag{24}$$

From this, we find that even the propagated diffraction field away from the restored image can be well explained within above scheme. We also note that the term of Eq. (23) calculated from the aliasing Fresnel factors appears a new form of point spread function with a spatial frequency of $\xi/\lambda z'$ where $z'$ is expressed as

$$z' = \frac{z - z_n}{z z_n}, \tag{25}$$

which might be due to a self-similarity of kernel function. In our previous work [9], we reported that the well diffracted fringe is observed as a function of a propagation distance away from the imaging plane when the diffraction field is simulated by using the enhanced-NA digital hologram. In the digital hologram with an aliased error from the sampling condition violation with respect to Fourier spectrum in Eq. (11), the rupture of diffraction fringe is observed.

As described previously, the resolution limit of restored image is decided by Eq. (6), whose relation is related to the NA of hologram. Above mathematical property also preserves the aperture size of digital hologram. That is, the property of Fresnel matrix is consistent with the conservation of hologram numerical aperture. The simple operation that truncates the hologram surely deteriorates image resolution [11]. We confirmed a deteriorated image in the truncated hologram with an overly aliased fringe. In addition, the hologram non-periodically undersampled worsens the image. We know that the spatial resolution of restored image is not governed by the maximum frequency of hologram fringe, but strongly depends on the NA of digital hologram.

## 3. Self-similarity of hologram fringe for undersampled digital Fresnel hologram

*3.1 Replication of concentric Fresnel zone by undersampling of digital Fresnel hologram*

For numerical simulation, the diffraction field can be written by the discrete form using the discrete Fourier transform ***DFT*** [1]:

$$U(m\Delta\xi, n\Delta\eta) = \frac{e^{ikz}}{i\lambda z}\exp\left\{i\frac{\pi}{\lambda z}\left[m^2\Delta\xi^2 + n^2\Delta\eta^2\right]\right\}$$
$$\times \boldsymbol{DFT}\left[U_0(p\Delta x, q\Delta y)\exp\left\{i\frac{\pi}{\lambda z}\left[p^2\Delta x^2 + q^2\Delta y^2\right]\right\}\right], \tag{26}$$

where the fields are digitized with steps $\Delta\xi$ and $\Delta\eta$ in the hologram plane, and $\Delta x$ and $\Delta y$ in the object plane. To analyze an aliasing effect through synthetic hologram, we calculate digital hologram about on-axis point sources located at various distances, in Fig. 1. The real component of the calculated diffraction field in the target plane is adopted as the hologram.

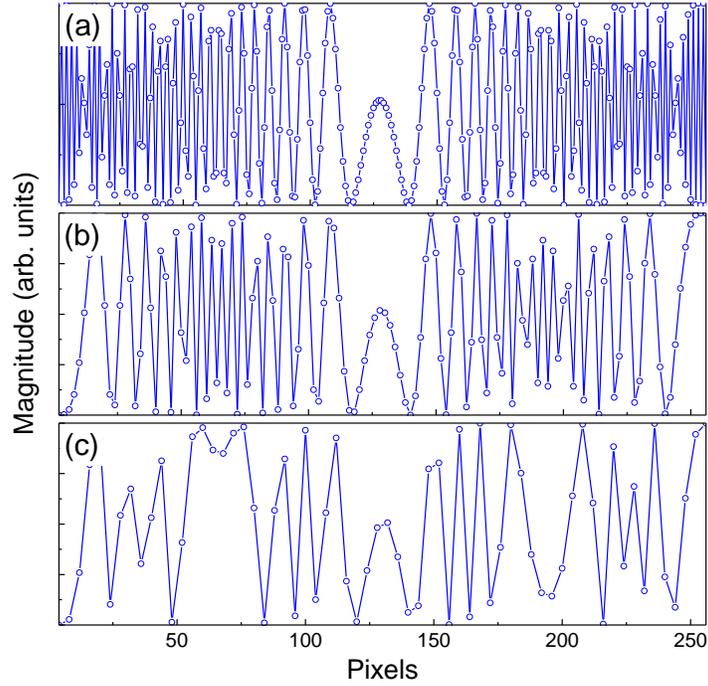

Fig. 2. Quadratic sinusoids of point spread function for a point source located at 30.8-mm distance. The curves are drawn along vertical line at a horizontal center of hologram and sampled by a sampling pitch of (a) 8 μm, (b) 16 μm, and (c) 32 μm.

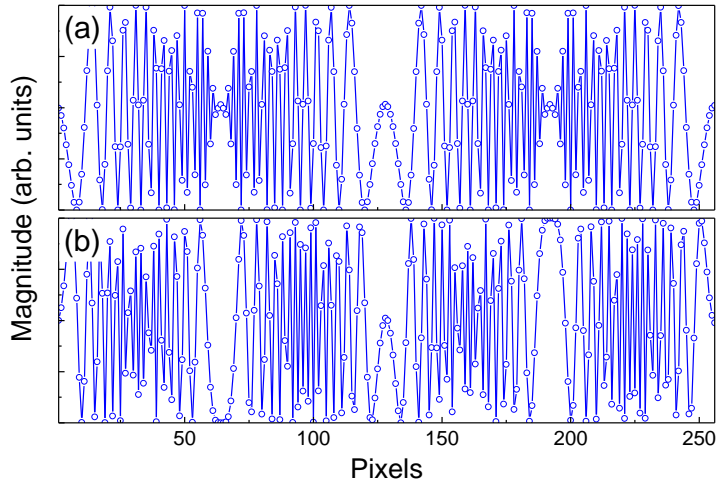

Fig. 3. Quadratic sinusoids of point spread function for point sources located at (a) 15.4-mm and (b) 7.7-mm distances.

The object and hologram has 256×256 pixels, where all the holograms have same pixel pitch of 8 μm. The coherent plane wave has 532-nm wavelength, and in this condition, the distance $z_0$ having 8-μm pixel pitch of the object is 30.8 mm. The relation of object and hologram resolutions with respect to a distance obeys Eq. (4). For convenience, we choose three points located at distances of a half and a quarter of $z_0$ as well as a distance of $z_0$. If the object field is put to be delta function $\delta(x, y)$ located on axis, **DFT** term becomes constant, and the hologram fringe is formed from the sampling of preceding quadratic phase term. The hologram fringe shape will be adaptively determined by the sampling interval, where the minimum distance to get correct sampling, $z_{\min} = N\varDelta\xi^2/\lambda$ .

In Fig. 2(a), the hologram synthesized using point source placed at *C* spot of $z_0$ distance shows the point spread function of a quadratic sinusoidal shape. Here, the function is drawn in one-dimensional $\xi$-coordinate. We know that as the resolution of hologram in the synthesis process is 8 μm, a complete sampling is expected, but we will show later that even this sampling is not sufficient. Especially, we find that the similar patterns of quadratic sinusoid are generated when the function is undersampled. The shape of replica patterns does not exactly coincide with each other. The number of the replicas grows in proportion to the increment of sampling pitch, where the curves of Figs. 2(b) and 2(c) are undersampled by pixel intervals of 16 μm and 32 μm, respectively.

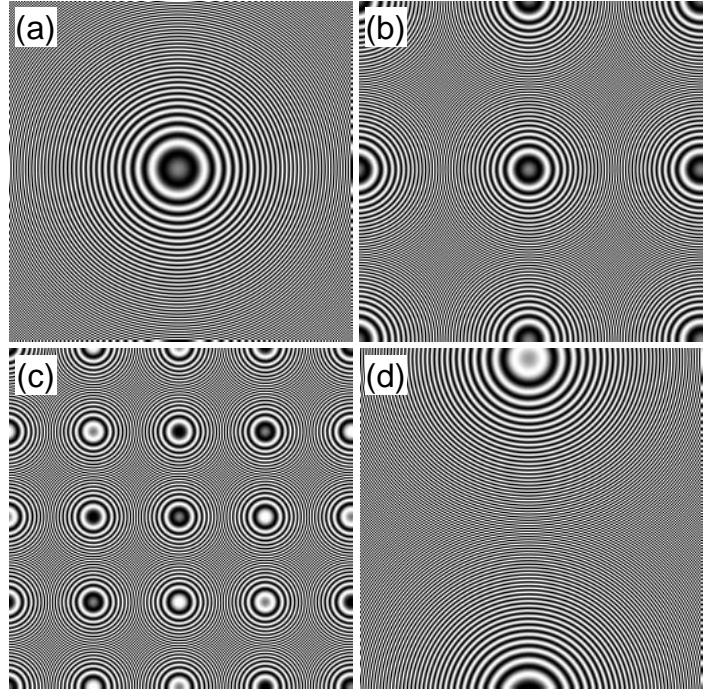

Fig. 4. Digital Fresnel holograms for on-axis point sources located at (a) 30.8-mm, (b) 15.4-mm, and (c) 7.7-mm distances. (d) Hologram fringe for off-axis point source at 30.8-mm is displayed.

We can observe naturally this behavior when synthesizing hologram by using point sources at closer spots, *B* and *A*, as depicted in Fig. 3. Since the maximum spatial frequency of digital hologram is decided from the Fresnel factor preceding the Fourier transform in Eq. (3), the value for the hologram synthesized at a distance 15.4 mm of $z_1$ is double in comparison to that for the hologram at $z_0$ distance, where 4-μm pixel sampling is required. Here, as the sampling pixel pitch is still 8 μm, the hologram fringe should be undersampled. We note that this geometrical situation fulfils well the Nyquist sampling criterion for the object field, based on Eq. (11), but the aliasing error of hologram fringe due to undersampling is created. Likewise, the hologram at a distance 7.7 mm of $z_2$ is overly undersampled owing to the requirement of 2-μm sampling interval. The density of quadratic sinusoids in Fig. 2(a), Fig. 3(a), and Fig. 3(b) are different because of difference in corresponding maximum frequencies. The diffraction angles $\Omega_0$, $\Omega_1$, and $\Omega_2$ with respect to the pixel resolution of object images are calculated to be about 3.8°, 7.6°, and 15.3°, respectively.

Figure 4 illustrates 2D hologram fringe of point sources, representing the sinusoidal Fresnel zone plate [23]. We can see a replication of similar zone pattern apparently. The number of replicas of zone plate increases by a square of scale. The four-fold undersampling forms zone plates of 16, in Fig. 4(c). Figure 4(d) is the hologram fringe for off-axis point source located at *E* spot in Fig. 1, where this spot is still placed within a diffraction area. Nevertheless, the replica zone plate by undersampling takes place, which is resulted from that a zone center placed at a boundary has two times the maximum frequency.

*3.2 Fractal character of point spread function*

We can observe that the peripheral zones between replicas come out blurry. To examine this phenomenon carefully, 2D hologram with 512×512 pixels is drawn in Fig. 5. The zoom-out operation of hologram in a digital display changes its shape due to undersampling. For clarity, the corresponding graph of quadratic sinusoid to the hologram fringe is drawn on, as depicted in inset of Fig. 5. It is certain that the sampled quadratic sinusoid has its similar pattern inherently. Particularly, a self-similar pattern emerges in the form of envelope curve in accordance with a scale. This is a fractal-like character having a self-similarity. In Fig. 2, the similar pattern of quadratic sinusoid emerges from the self-similar pattern of the envelope curve when being undersampled. We find that the sampling operation to the point spread function induces its fractal structure, which plays a role in making an aliasing fringe in digital hologram.

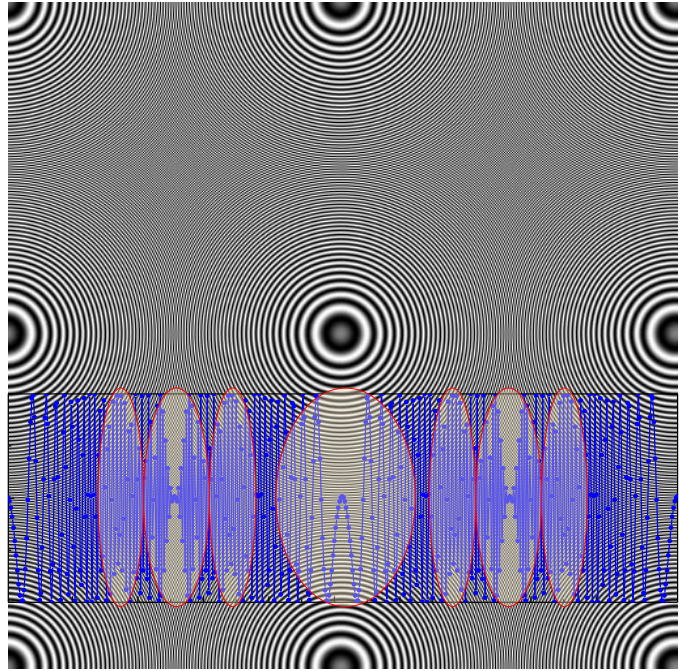

Fig. 5. Digital Fresnel hologram for on-axis point source at 30.8-mm distance. The hologram is synthesized with object of 512×512 pixels. When the hologram is zoomed out in a digital display, a similar pattern of Fresnel zone is appeared constantly due to undersampling. The inset graph reveals a self-similar envelope pattern of quadratic sinusoid.

This fractal-like structure shows an extraordinary behavior. That is to say, as a conventional fractal image continues to be magnified by a scaling parameter, its similar pattern shows up constantly [24]. The envelope shape of zone pattern showing a fractal-like structure seems to correspond to this condition. However, an apparent zone plate is generated only by undersampling process with an increased pixel interval. This can be viewed as the zoom-out process, but this property is strictly irrelevant to the negative Fractal dimension [25], because as confirmed in Fig. 2, physical size of similar patterns is constant. The fractal dimension for hologram fringe could be calculated by using the box-counting method. However, because the self-similar curves are in the form of the envelope, the box-counting technique would be cautiously treated.

The fractal structure in nature has been known to be in the nonlinear dynamics [26]. Therefore, it is interesting that the point spread function derived from the linear differential equation for optical wave propagation has a fractal-like structure. As previously described, the several fractals have been found even in linear optical elements [19,27,28]. The local foci of a binary diffraction lens reveal the fractal distribution in the axial direction, and self-images generated from the linear grating element have a fractal structure. Since this property arises by the interaction of point spread function and discrete element, the internal structure of the optical kernel function could closely relate to these fractal behaviors of the diffraction beam, and thus, a quantitative study is required for detailed understanding.

## 4. Numerical analysis of aliasing phenomenon and image recovery for digital Fresnel hologram with enhanced numerical aperture

### 4.1 Aliasing phenomenon of digital Fresnel hologram synthesized with finite object

We confirmed that even digital holograms suffering a severe aliased fringe such as Fig. 4(c) recover the original point object through inverse Fresnel transform. This aliased hologram could play a role in multifocusing Fresnel lens, i.e. the replica zones could produce the replica point images which match with the high-order images generated from the pixel pitch [16,17]. The coincidence between the replica images and high-order images is ultimately resulted from the self-similarity of the optical kernel function.

The point object is only mathematical concept because physical object should have a finite extent. Physical objects can be considered as an aggregation of point objects. Thus, the object having a finite size will reveal a size effect for optical diffraction. For the further study, we simulate the hologram synthesis and its image reconstruction by using circular objects with a finite size and 'HOLO' letter object. The simulation specifications are put to be those of four-fold undersampling in Fig. 4(c). Figure 6 shows the hologram fringe property for synthesized holograms.

We write the Fresnel diffraction from the object with a finite extent $l$ in the convolution form of the object field and point spread function:

$$U(\xi,\eta) = U_0(\xi,\eta)\text{rect}\left(\frac{\xi}{l},\frac{\eta}{l}\right) * h(\xi,\eta). \tag{27}$$

Above equation is rewritten by

$$U(\xi,\eta) = \frac{e^{ikz}}{i\lambda z}\exp\left[\frac{i\pi}{\lambda z}(\xi^2+\eta^2)\right]\textbf{\textit{FT}}\{U_0(\xi,\eta)h(\xi,\eta)\} * \text{sinc}\left(\frac{\pi\xi l}{\lambda z}\right)\text{sinc}\left(\frac{\pi\eta l}{\lambda z}\right). \tag{28}$$

The Fresnel number $N_F = l^2/\lambda z$ can be a criterion to discriminate the far-field and near-field diffractions. In case the object size is close to the point object, the Fraunhofer diffraction evolves, where in Eq. (27), the Fourier transform term of a rectangular function becomes a modulating sinc function. As the object size increases, the maximum peak width $\lambda z/l$ of sinc function decreases. As shown in Fig. 6, the size effect of diffraction becomes stronger with increasing a radius of circular object. That is to say, the diffracted wave from object with a large extent propagates more straight forward, which is a typical property showing in the near-field diffraction at a large $N_F$. This behavior does not mean the change in the width of propagated diffraction spectrum, because the whole range of diffracted field is fundamentally determined by the object resolution according to Eqs. (4) and (7).

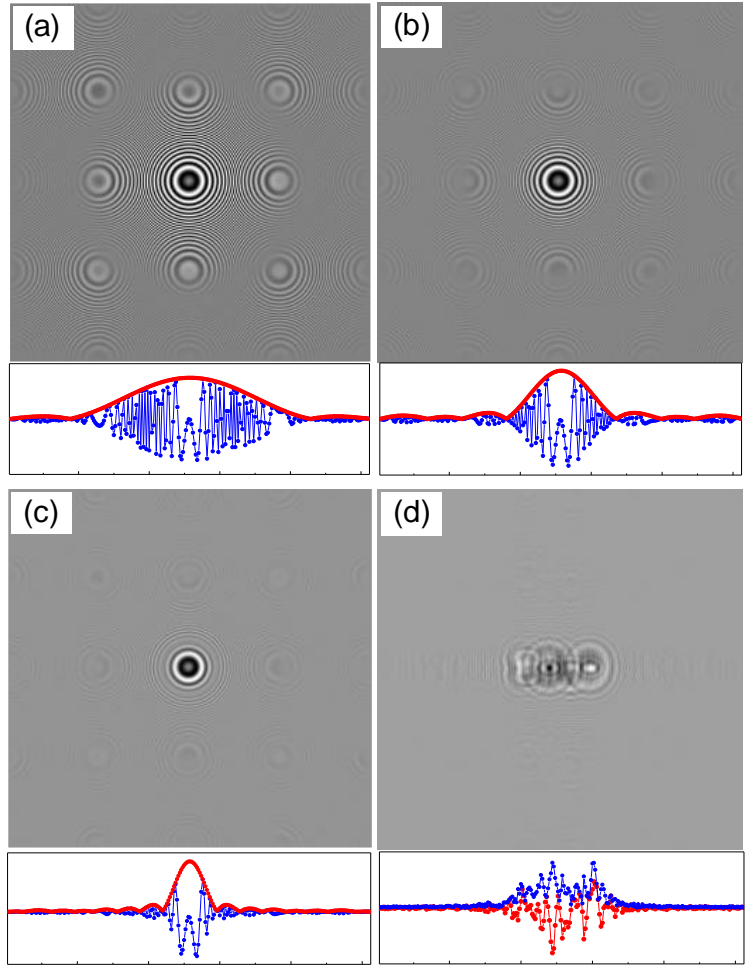

Fig. 6. Digital Fresnel holograms synthesized using circular objects with a radius of (a) 4 μm, (b) 8 μm, and (c) 16 μm, and (d) 'HOLO' letter object. Below graphs indicate one-dimensional hologram fringe and its intensity profile in the horizontal direction at the center of image.

The digital holograms synthesized by using circular objects with a radius of 4 μm, 8 μm, and 16 μm are displayed in Figs. 6(a)-6(c). We find that as with hologram fringe of point source, the replica of concentric fringe is formed. However, when object size is large, the modulus of replica fringe is considerably suppressed by a modulating sinc function. Likewise, the hologram fringe of letter object reveals the suppressed replicas of interferogram positioned at a center, in Fig. 6(d). The suppressed replicas of interfergrams cannot generate replica images other than the replica hologram zones arising from point object. Meanwhile, the whole area of digital hologram contributes to the retrieval of object image. From this property, we find that the digital hologram synthesized at a closer distance than $z_0$ have an enhanced numerical aperture, which results in the recovery of object image with higher resolution.

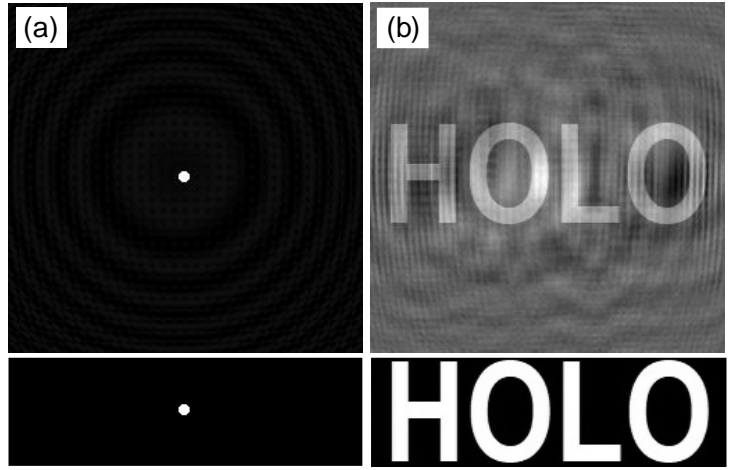

Fig. 7. Restored images from real-valued digital holograms for (a) circular object with a radius of 8 μm and (b) letter object. Below figures display the restored images from complex holograms.

*4.2 Numerical simulation of image recovery for digital Fresnel hologram*

Figure 7(a) is the restored image of circular object with a radius of 8 μm by using a real-valued hologram. Since the resolution of restored image is 2 μm, the total image size becomes 512 μm. We can clearly distinguish the object image from its conjugate term. We know that in in-line holographic system, only the real-valued or imaginary-valued hologram makes it difficult to analyze the aliasing effect owing to an overlap of twin image. Below image is the restored image from complex hologram, which shows a most complete retrieval. The restored image for letter object is displayed in Fig. 7(b). The real-valued hologram reconstructs the discriminated original image clearly.

Based on the analysis in Subsection 4.1, we can extract that the digital hologram is effectively synthesized using the object being extended to the deviating area from a diffraction zone by pixel pitch [9]. This makes it possible to recover the reconstructed image shrinkage suffered in enhanced-NA hologram. As depicted in Fig. 8, we consider the extended object added by a rectangular object outside of original letter object, which is located at the $z_1$ distance of 15.4 mm in Fig. 1. The extended object with 512×512 pixels of a 4-μm resolution has the same size as 2048-μm of the hologram. The angle value angle $\Omega_1$ with respect to object pixel pitch 4 μm is 7.6°, and thus, the calculated diffraction field places in the diffraction zone. The digital hologram is calculated from the Riemann integral of the Rayleigh-Sommerfeld diffraction formula to control arbitrary the input and output sizes. To fulfil the Nyquist sampling criterion enough, the digital hologram is calculated through a two-fold upsampling of object field. The aliasing error due to a self-similarity of point spread function will be inhibited in hologram fringe although it is not apparent to the naked eye.

The reconstructed image by using complex hologram in Fig. 8(a) reveals the overlapping of high-order aliasing images because of a finite diffraction zone of 8-μm pixel pitch, where the diffraction area of 1024 μm is a half of the object field-view of 2048 μm. This high-order aliasing can be largely suppressed by a diffraction zone enlargement from two times upsampling of hologram fringe, as displayed in Fig. 8(c). However, it is not completely removed, because although the diffraction angle of pixel pitch is double, the configuration of replicated Fourier spectrums is kept. Here, only the quadratic phase term in parenthesis of the Fourier transform is affected from the extension of diffraction angle.

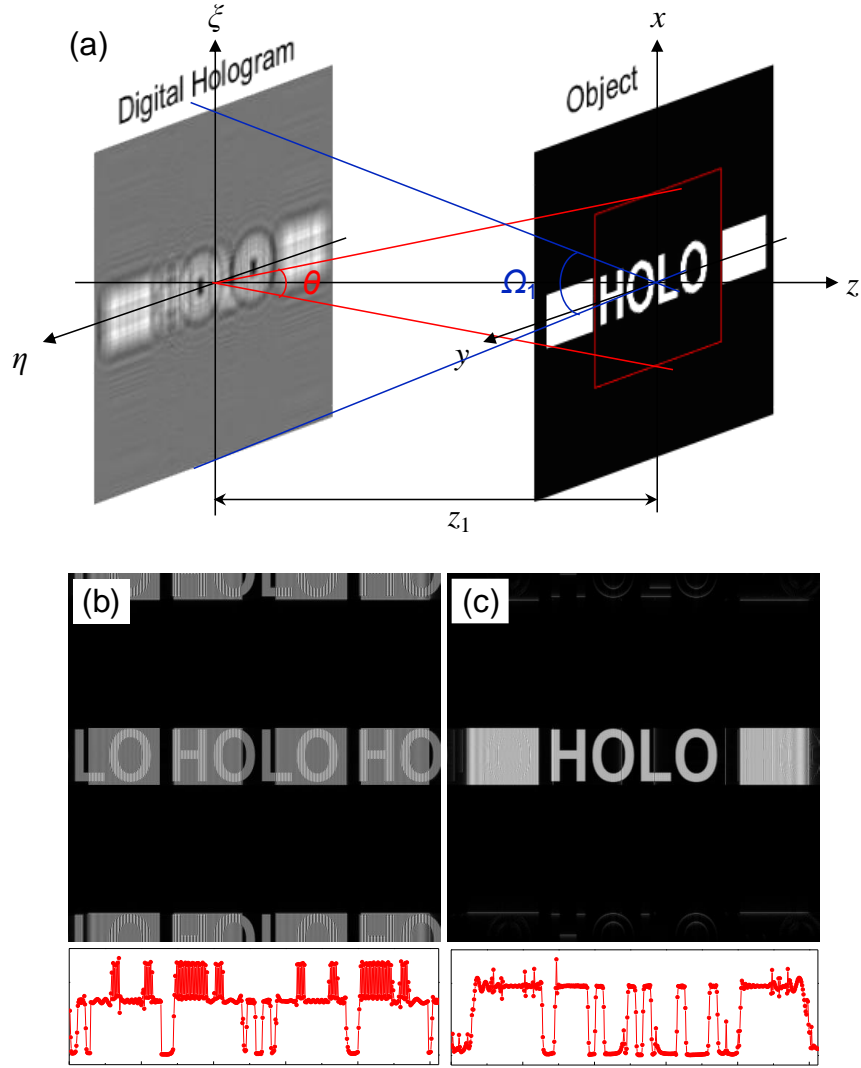

Fig. 8. Simulation results of image reconstruction for (a) the hologram synthesized using extended object: (b) Restored image from original hologram, and (c) restored image from two-fold upsampled hologram. Below graphs indicate the intensity profile in the horizontal direction at the center of image.

## 5. Discussions

We observed optically the well-reconstructed image from the digital hologram with enhanced-NA, in our previous work [9]. Especially, we elucidated that the active diffraction angle is determined by the hologram numerical aperture, and the viewing-angle $\Omega$ of holographic image is written by

$$\Omega = 2\sin^{-1}\left(\frac{N\Delta\xi}{2z}\right). \tag{29}$$

This property allows us to develop the holographic display with a wide-viewing angle. Present spatial light modulator to load the digital hologram has a pixel pitch with several micrometer scale, and thus, the viewing-angle of holographic image is limited to several degrees, which is known as the critical obstacle for realizing the holographic display. According to our scheme, the computed generated hologram with the enhanced-NA loaded on the spatial light modulator can reconstruct optically the image with a wide angle compared to that by the modulator pixel pitch, and without the shrinkage of image.

In the hologram acquisition system, the digital hologram could be adaptively acquired in accordance with a pixel pitch of digitized sensor. Since a real object is continuous, it does not have the limitation of diffraction angle. Conversely, the spectral window of captured diffraction field defines the image resolution depending on a numerical aperture. If we consider the real object as a collection of the point object appeared as a delta function, the spatial information is encoded in the shape of fringe pattern. The fringe pattern will depend on the lateral size of sensor and acquisition distance. The hologram fringe is undersampled by a finite pixel pitch when it is captured at a closer distance than $z_0 = N \varDelta \xi^2 / \lambda$, which leads to an aliasing error of hologram fringe. Our result reveals that even digital hologram suffering a severe aliasing error can restore the original almost completely. In this acquisition system the object reduction is inevitable, but the limitation of object size can be also overcome by enlargement of diffraction zone for digital hologram. After all, our work shows the possibility for retrieving the high resolution image with a sufficient extent irrespective of the degree of aliasing error of hologram fringe.

## 6. Conclusions

We elucidate that optical kernel function reveals a fractal-like behavior when being sampled. A self-similarity of concentric Fresnel zone becomes a crucial mechanism in inducing an aliasing fringe for digital hologram with the enhanced numerical aperture and restoring a holographic image stably. The enhanced-NA digital hologram acquired at a closer distance from the object could have an aliased fringe due to a finite pixel size of digitized device. The original image can be almost completely retrieved even under this severe aliasing phenomenon. We find that the spatial resolution of restored image is determined by a numerical aperture of digital hologram other than the maximum frequency of hologram fringe. This result makes it possible to apply for obtaining the high-resolution holographic image in holographic display and in optical hologram acquisition systems.

## Acknowledgements

This work was partially supported by Institute for Information & Communications Technology Promotion (IITP) grant funded by the Korea government (MSIP) (2017-0-00049)